\documentclass[aps,twocolumn,10pt,nofootinbib,superscriptaddress,hyperref]{revtex4}
\usepackage{graphicx}
\usepackage{epsfig}
\usepackage{color}
\usepackage{amsmath}
\usepackage{amsfonts}
\usepackage{amssymb}
\usepackage{cancel}
\usepackage{url}

\begin{document}
\pagestyle{plain} 
\title{\bf Naturally Light Sterile Neutrinos from Theory of R-parity}
\author{Dilip Kumar Ghosh}
\affiliation{Department of Theoretical Physics, Indian Association for the Cultivation of Science, 
2A\&2B Raja S.C.\,Mullick Road, Kolkata 700\,032, India}
\author{Goran Senjanovi\'c}
\affiliation{High-Energy Section, Abdus Salam International Center for Theoretical Physics,\\ Strada Costiera 11, I-34014 Trieste, Italy}
\author{Yue Zhang}
\affiliation{High-Energy Section, Abdus Salam International Center for Theoretical Physics,\\ Strada Costiera 11, I-34014 Trieste, Italy}
\date{\today}
\begin{abstract}
The fate of R-parity is one of the central issues in the minimal supersymmetric standard model (MSSM). Gauged $B-L$ symmetry provides a natural
framework for addressing this question. Recently, it was pointed out that the minimal such theory does not need any additional Higgs
if the $B-L$ breaking is achieved through the VEVs of right-handed sneutrinos, which ties the new physics scale to the scale of the MSSM. 
We show here that this 
immediately leads to an important prediction of two light sterile neutrinos, which can play a significant
role in the BBN and neutrino oscillations. We also discuss some new relevant phenomenology for the LHC, in the context of the minimal supersymmetric left-right symmetric theory which provides a natural setting for the gauged $B-L$
symmetry.
\end{abstract}
\maketitle 

\section{Introduction}

Over the decades, low energy supersymmetry became the central extension of the standard model (SM). 
Its attributes are well known: stability of the gauge hierarchy, gauge coupling unification \cite{susyunif} and the radiative Higgs
mechanism \cite{Inoue:1982pi}. However, supersymmetry allows for new interactions which break the $B$ and $L$ accidental symmetries of the SM.
The stability of the proton and the associated phenomena become a mystery.  Due to non-renormalization theorems,
one can simply make the new interactions (or a subset of them) vanish. Only if all of them end up vanishing, one gets the well-known
R-parity symmetry, under which ordinary particles are even and super-partners are odd. R-parity is 
equivalent to matter parity
\begin{equation}
M_p = (-1)^{3(B-L)} \ .
\end{equation}
A priori, there is no way of knowing whether or not R-parity is a good symmetry.
Gauging $B-L$ symmetry sheds light~\cite{Mohapatra:1986su} on the fate of this otherwise {\it ad hoc} symmetry. This can be achieved at the expense
of introducing right-handed neutrinos, which is the simplest way of generating neutrinos masses. 
The right-handed neutrinos are particularly natural in the context of left-right (LR) symmetric theories~\cite{Pati:1974yy}, which provide a
possible understanding for parity violation in nature. Furthermore, they contain automatically gauged $B-L$ symmetry.
In short, these theories offer a unified picture of three fundamental issues: the origin of parity violation, neutrino mass and
the fate of R-parity.

Recently new progress was made in the construction of these theories, by achieving the breakdown of $B-L$ and/or LR symmetries through the RH sneutrino VEVs only~\cite{Barger:2008wn, FS}, which simultaneously breaks R-parity.
These are arguably minimal such models and furthermore, they tie automatically the gauge symmetry breaking scale to the scale of supersymmetry breaking soft terms. This provides a great boost towards the detectability of this class of theories at the large hadron collider (LHC).

We show here, that these theories have an additional important feature: the necessary existence of two light (sub-eV) sterile neutrinos
which may play a crucial role in the anomalies of neutrino oscillations. Furthermore, due to the new
gauge interactions, they are equally abundant as the left-handed ones at the time of big bang nucleosynthesis (BBN). This is the central finding of this Letter. The latest analysis of BBN seems to favor such degrees of freedom~\cite{Hamann:2010bk}.  We also discuss some new associated phenomenology
for the LHC in the context of the LR symmetric version of the theory.

Before turning to the description of the model, let us clarify the essence of our result. First of all,
what role does supersymmetry play? After all, gauging $B-L$ is interesting in its own right and it is worth discussing
what happens without supersymmetry. Obviously, the neutrino spectrum depends crucially on what Higgs one chooses in order 
to break $B-L$. In the seesaw picture~\cite{theseesaw}, one opts for the $B-L=2$ SM singlet field. One may instead prefer the $B-L=1$ Higgs field, 
in which case at the renormalizable level, the RH neutrinos do not get Majorana masses and one is left with 3 Dirac neutrinos. This was discussed in Ref.~\cite{Barger:2003zh}. The modern point of view, though, is not to ignore higher dimensional operators, as illustrated by the famous $d=5$ Weinberg operator in the SM~\cite{Weinberg:1979sa}. The analogous operator in this case
implies the RH neutrino masses on the order of $M_{BL}^2/M$, where $M$ is the scale of new physics above $M_{BL}$.
Both of these scales are arbitrary and no prediction can be made.
One could also choose a large $B-L$ charge ($\geq3$) for the SM singlet Higgs, in which case, RH neutrinos would remain massless and again one would end up with three Dirac neutrinos. In short, the resulting physics depends on the arbitrary
model building and/or assumptions about higher dimensional operators, with no information about mass scales.

On the other hand, as we argued, the supersymmetric $B-L$ theory in its minimal form relates the scale of $B-L$ breaking
to the scale of low energy supersymmetry, and leads automatically to two extra sterile neutrinos. 
In what follows, we show now how this takes place.

\section{The Supersymmetric $U(1)'$ model}

The minimal model~\cite{Barger:2008wn} of gauged $B-L$ symmetry is based on the gauge group
\footnote[1]{Strictly speaking, one can start with two $U(1)$ symmetries, which are arbitrary linear combinations of $B-L$ and hypercharge. The difference lies in the couplings of $Z'$ boson to the SM fermions.}
 $G_{B-L}=SU(3)_c \times SU(2)_L\times U(1)_Y \times U(1)_{B-L}$. The matter assignment is given by 
\begin{equation}\label{fermions}
Q = \left( \begin{array}{c} u \\ d \end{array} \right), \ \  L =  \left( \begin{array}{c} \nu \\ e \end{array} \right), \ \ u^c, \ \ d^c, \ \ e^c, \ \ N^c, \
\end{equation}
where beyond the usual SM particles, one needs (anti) RH neutrinos, one per generation, in order to cancel the $B-L$ anomaly.
The most general gauge-invariant superpotential becomes then (we ignore the quark sector being irrelevant for what follows)
\begin{equation}\label{3}
W = Y_\nu^D L H_u N^c + Y_e L H_d e^c + \mu H_u H_d , 
\end{equation}
where $H_u, H_d$ are the MSSM Higgs doublets. 
In this minimal model, one
does not introduce any new Higgs multiplets so that $B-L$ symmetry must be broken by the non-vanishing VEV of the RH sneutrino field $\langle\widetilde N^c\rangle$.
This requires a tachyonic soft mass for the $\widetilde N^c$ field.

The most appealing feature of this approach is that the scale of $B-L$ breaking is tied to the scale of soft supersymmetric terms, and thus if the MSSM
is to be observed at LHC, so would be the new gauge interactions. Another important feature is the prediction of light sterile neutrinos, as we show now.

\subsection{Neutrino Masses at Tree Level}
The essential point is one RH neutrino becomes massive at the first stage of symmetry breaking. One is always free to rotate the RH sneutrino VEVs
into the direction $\langle\widetilde N_1^c\rangle$. In turn one gets
\begin{equation}\label{mass}
M_{N^c_1,\widetilde Z_{BL}} = \left( \begin{array}{cc} 0 & M_{Z'} \\ M_{Z'} & m_{1/2} \end{array} \right),\ \ M_{N^c_2} = M_{N^c_3} = 0, \ 
\end{equation}
where $M_{Z'}=g_{BL}\langle\widetilde N_1^c\rangle$ is the mass of the $Z'$ boson, $\widetilde Z_{BL}$ is the gaugino of $U(1)_{B-L}$ with generic Majorana soft mass $m_{1/2}$ and $g_{BL}$ is the $B-L$ gauge coupling.
For simplicity, from now onwards, we will assume roughly one scale of soft supersymmetry breaking terms, which is denoted as $M_{SUSY}$, thus $m_{1/2}\simeq M_{SUSY}$. 
 
After the second stage of symmetry breaking where $H_u, H_d$ gets their VEVs, from Eq.\,(\ref{3}), one gets Dirac neutrino mass terms.
One linear combination of LH neutrinos, called $\nu_1$ hereafter, get a Majorana mass $m_{\nu_1} \simeq (Y_
\nu^{D})_{11}^2 \langle H_u\rangle^2 / M_{N_1^c}$ through the seesaw mechanism, when $N^c_1$ gets integrated out 
\footnote[2]{Notice there is another contribution due to the LH neutrino-Higgsino-neutral gaugino mixing, which gives Majorana mass to the same linear combination $\nu_1$.}.
To implement TeV seesaw, we need $(Y_\nu^D)_{11}\lesssim 10^{-6}$.
In principle, the left-handed sneutrino $\widetilde\nu$ can also get a VEV, which has to be very small. The point is that it mixes LH neutrinos with neutralinos. 
A different linear combination of LH neutrinos, called $\nu_2$ hereafter, gets a Majorana mass $m_{\nu_2}\propto \langle \widetilde\nu\rangle^2/m_{1/2}$, when neutralinos are integrated out. 
Neutrino oscillations together with beta decay require the mass eigenstates to lie below eV.
For $m_{1/2}\simeq100\,{\rm GeV}-{\rm 1TeV}$, from $m_{\nu}\lesssim$ eV, one gets $\langle \widetilde\nu\rangle\lesssim 300\,{\rm KeV-1 MeV}$. 

In general, the five light neutrino states $(\nu_1, \nu_2, \nu_3, N^c_2, N^c_3)$ will mix and produce the following mass matrix at renormalizable level
\begin{equation}
M_{\rm light} = \left( \begin{array}{ccccc}
m_{\nu_1} & 0 & 0 & m_D^1 & m_D^2 \\
0 & m_{\nu_2} & 0 & m_D^3 & m_D^4 \\
0 & 0 & 0 & m_D^5 & m_D^6 \\
m_D^1 & m_D^3 & m_D^5 & 0 & 0 \\
m_D^2 & m_D^4 & m_D^6 & 0 & 0 \\
\end{array} \right), 
\end{equation}
where the state $\nu_3$ is orthogonal to $\nu_1, \nu_2$ states and $m_D^i\simeq Y^D_\nu \langle H_u\rangle$ are the Dirac masses in this basis. At tree-level, there are five Majorana light states. Their masses are not predicted and must be inferred from experiment. 
There is the cosmological bound from the WMAP on the sum of all light neutrino masses, again on the order of eV.
This implies the corresponding $(Y_\nu^D)_{ij} \lesssim 10^{-11}$ for $i=1,2,3$ and $j=2,3$. 

\bigskip
The extra two light states may be welcome~\cite{Akhmedov:2010vy}, in view of the oscillation anomalies in LSND, MiniBooNE and MINOS~\cite{LSND, miniboone, minos}. 
Furthermore, the case is being made precisely for two more light fermionic states, in order to have successful BBN~\cite{Hamann:2010bk}.
These states by definition, should be roughly equally abundant as the usual three active neutrino in the SM.
This is guaranteed due to the $B-L$ gauge interactions, which
keep the light states $N^c_2, N^c_3$ in thermal equilibrium until the temperature falls down to about 10\,MeV. 
In principle Yukawa coupling could also do the job~\cite{Acero:2008rh},
which would require a quantitative analysis, not needed here.

\subsection{Higher Dimensional Operators}

How robust are these predictions? What happens if one introduces higher dimensional operators?
The gauge singlet combinations which are 
building blocks for such operators are 
\begin{equation}
L H_u N^c, \ \ \ H_u H_d, \ \ \ L H_d^\dag N^c, \ \ \ N^{c\dag} N^c \ .
\end{equation}
The first two appear in the superpotential at tree level and thus offer nothing new. In this sense they can be classified as
trivial operators, since they are basically renormalizing the already present couplings and masses. 
The typical operators that we have in mind can be represented by the following F-type term
\begin{equation}
O_F = (LH_uN^c) \frac{H_u H_d}{M^2} \ .
\end{equation}
This would induce a Dirac mass term of the order of $m_D\simeq M_W^3/M^2$. Clearly, either the scale of new physics
is large enough, $M\gtrsim3\times 10^7\,$GeV or it requires a small Wilson coefficient.

Similarly, the third term in Eq.~(6) also gives a Dirac mass term from K\"ahler potential, $m_D=
\langle F_{H_d}\rangle^\dag/M = \mu \langle H_u\rangle/M \simeq M_{W}M_{BL}/M$m which now requires
$M\gtrsim 10^{14}\,$GeV. This is in complete analogy with the situation in the SM, where $d=5$ Weinberg
operator requires a large scale or a small coupling.

Notice this is nothing new. In the SM, the tree level Yukawa couplings are similarly affected by $(H^\dag H/M^2)^n$ Wilson-like terms. Small Yukawa couplings as above imply large $M$ or small coefficients. This is the cornerstone of the SM flavor structure. For this reason we do not discuss such operators anymore.

More interesting is the fourth type gauge invariant operator in Eq~(6), which affects the tree-level K\"{a}hler potential. 
Together with $H_u H_d$, it produces a general class of operators relevant for RH neutrino Majorana masses~
\footnote[3]{In principle, the operators  contain additional factors $(H_u^\dag H_u)^m (H_d^\dag H_d)^n$, but they clearly lead to same physics.}
\begin{equation}
O_D^{(p, q)} = \frac{ N_1^{c\dag} N_1^{c\dag} N_a^c N_b^c }{M^2} \left( \frac{H_uH_d}{M^2} \right)^p \left( \frac{H_u^\dagger H_d^\dagger}{M^2} \right)^{q} \ ,
\end{equation}
where $a, b = 2, 3$. The case $p=q=0$ leads to the sterile neutrino mass matrix (corresponding to the lower-right $2\times 2$ submatrix of Eq.~(5))
\begin{equation}
m^{ab}_{N^c} (0,0) = \frac{\langle \widetilde N_1^{c\dagger} \rangle \langle F_{N_1^c}^\dagger \rangle}{M^2} \simeq Y_\nu^D \frac{\langle \widetilde N_1^{c\dag} \rangle \langle H_u \rangle \langle \widetilde\nu \rangle}{M^2} \leq 10^{-3} {\rm eV} ,
\end{equation}
for $M\gtrsim 10 M_{Z'}$. The reason this comes out negligible is the double suppression of small $Y_\nu^D$ and small $\langle \widetilde\nu \rangle$.

Ironically, the leading contribution comes from the next order operator with $p=0$, $q=1$, which now leads to
\begin{equation}
m_{N^c}^{ab}(0,1) = \frac{\langle \widetilde N_1^{c\dagger} \rangle^2 \langle H_u \rangle \langle F_{H_d}^\dagger \rangle}{M^4} \simeq \frac{\mu \langle \widetilde N_1^{c\dag} \rangle^2 \langle H_u \rangle^2 }{M^4}\lesssim{\rm MeV}\ .
\end{equation}
for $M\gtrsim 10 M_{Z'}$. It is much larger than $m^{ab}_{N} (0,0)$ since $F_{H_d}\gg F_{N_1^c}$. 

If these sterile neutrinos $N_{2,3}$ are to have appreciable masses, they must decay. 
The lighter of the two (called $N_L$ hereafter) decays through the dipole operator, which is generated at the loop level.
\begin{equation}
\frac{e}{16\pi^2} \frac{m_D}{M_W^2} \bar \nu \sigma_{\mu\nu} N_L F^{\mu\nu} \ ,
\end{equation}
while the heavier one (called $N_H$) can also decay via the higher dimensional operator from the K\"ahler
\begin{equation}
\frac{ L^\dag LN_L^\dag N_H}{M^2}\ ,
\end{equation}
It is easy to verify that both lifetimes turn to be longer than $10^{12}\,$sec, for $m_{N_a} \lesssim\,$MeV, or
cosmologically stable for  $m_{N_a} \lesssim 100\,$KeV. Since $N_L$, $N_H$ are roughly as abundant as the photons,
unless being very light, they would quickly dominate the universe
when the photon temperature drops below their mass.
This would run into conflict with observations on BBN, CMB and large-scale structure.
Recent studies of the constraints from structure formation find $m_N^{ab}\lesssim 1\,$eV~\cite{Krauss:2010xg}, and 
in turn $M\gtrsim3\times 10^3M_W$. 

\bigskip
What about the loops? They are actually very tiny if nonzero,
for they must go through the Dirac couplings $Y_\nu^D$, which is the only source of the breaking of flavor symmetry in the neutrino sector. 

\bigskip
In summary, should the future confirm the need for sterile neutrino masses in the eV or sub-eV region, the scale $M$ of new physics
beyond $B-L$ would have to be fairly low, on the order of $10^5\,$GeV. This is good news since any oasis in the desert 
is welcome.

\section{Left-right symmetric version}
The minimal supersymmetric LR model (which we call the FS model~\cite{FS} here and below) is based on the gauge group $G_{LR}=SU(3)_c\times SU(2)_L\times SU(2)_R\times U(1)_{B-L}$. Besides the $Q$ and $L$ doublets, one has the singlet fermions of Eq.~(\ref{fermions}) as the $SU(2)_R$ doublets 
\begin{equation}
Q^c = \left( \begin{array}{c} u^c \\ d^c \end{array} \right), \ \  L^c =  \left( \begin{array}{c} N^c \\ e^c \end{array} \right)\ .
\end{equation}
Now $H_u$ and $H_d$ form a bidoublet $\Phi$ under $SU(2)_L\times SU(2)_R$ and as before no new Higgs field is introduced. Again R-parity is a gauge symmetry and is spontaneously broken by the VEV of RH sneutrino. 
This requires the asymmetric soft terms for left and right sleptons, as assumed in~\cite{FS}. 
Strictly speaking, this is not a left-right symmetric theory although the breaking is soft.
In principle, this soft breaking can emerge from the D-term VEV of a parity odd singlet (partially) responsible for supersymmetry breaking in the hidden sector.

The characteristic of the model is the prediction for the ratio of heavy neutral and charge gauge bosons
\begin{equation}
\frac{M_{Z'}}{M_{W_R}}=\frac{g_R/g_L}{\sqrt{(g_R/g_L)^2 -\tan^2\theta_W}} > 1 \ ,
\end{equation}
where $\theta_W$ is the Weinberg angle. Besides the usual upper limit on $g_R$ from the perturbativity requirement, it is also easy to see there is a lower limit, $g_R>g_L\tan\theta_W$.
When $g_L\approx g_R$, one gets $M_{Z'}/M_{W_R}= \cos\theta_W/\sqrt{\cos2\theta_W}\approx 1.2$.
It should be contrasted with the popular version of the theory based on the renormalizable seesaw mechanism. In the latter case, one utilizes the triplet Higgs to break the $SU(2)_R$ symmetry and give mass to all $N^c$, and one gets $M_{Z'}/M_{W_R}\approx 1.7$ for $g_L\approx g_R$.

\subsection{Neutrino Masses}

What about the neutrino mass spectrum in this theory?
In the same manner as in the $U(1)'$ model, one ends up with only one heavy RH neutrino at tree level.
The point is that we only added RH charged current interactions without modifying the neutral fermion mass matrix structure.
Regarding the higher dimensional operators, the conclusion goes through as before, which again implies two light sterile neutrinos.
The only difference lies in the loop corrections to the sterile neutrino masses. There is a new contribution 
shown in Fig.~1.
 
The point is that in the LR model, there are bidoublets, and so the $y_\tau$ coupling might enter the above loops. The most general superpotential takes the form 
\begin{eqnarray}
W = Y_\ell L \Phi_1 L^c + Y_\nu^D L \Phi_2 L^c + \mu_{ij} {\rm Tr} [\Phi_i\Phi_j] \ ,
\end{eqnarray}
where $\Phi_i=(H_u, H_d)_i$, $i,j=1,2$, with the most general VEVs
\begin{eqnarray}
\langle \Phi_{1} \rangle = \left( \begin{array}{cc} 0 & 0 \\ 0 & \kappa_1 \end{array} \right), \ \ \ 
\langle \Phi_{2} \rangle = \left( \begin{array}{cc} \kappa_2' & 0 \\ 0 & \kappa_2 \end{array} \right),
\end{eqnarray}
In this case, the Dirac neutrino mass is still proportional to $Y_\nu^D$, with $(y_\nu^D)_{11} \sim 10^{-6}$, $(y_\nu^D)_{ij} \lesssim 10^{-11}, (i, j = 2, 3)$, similar to the B-L model.
The main contribution to charged lepton mass is from the first Yukawa $Y_\ell={\rm diag}(y_e, y_\mu, y_\tau)$ and the second term is only a small correction.
A crucial point to note is that $H_{u1}$ has no VEV but large Yukawa couplings $y_\tau$ to the leptons/neutrinos, while $H_{u2}$ could have large VEV but small Yukawa $Y_\nu^D$.
The contribution from Fig.~1 is given by
\begin{eqnarray}
(m_{N^c})_{ab} \sim \frac{g y_\tau^4}{(16\pi^2)^3} m_{1/2} \simeq 10^{-14} m_{1/2} \tan^4\beta \ ,
\end{eqnarray}
which is sub-eV as long as the model lies in the low $\tan\beta = \kappa_2'/\kappa_1$ regime.

\begin{figure}[h!]
\begin{center}
\includegraphics[width=7cm]{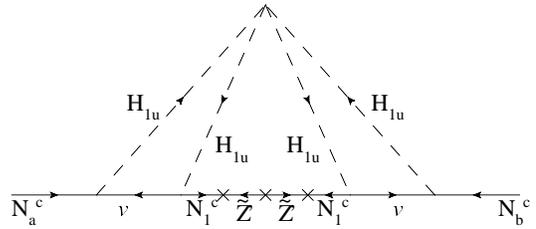}
\caption{The Cheops diagram: potentially large radiative correction to the sterile neutrino masses, where $a,b=2,3$.}\label{fig1}
\end{center}
\end{figure}

\subsection{Collider Signatures}
The virtue of the LR model, is the possible lepton number violation at the colliders through the production of $W_R$. 
This happen when the RH neutrino picks up a large Majorana mass, and the end result is same-sign dileptons~\cite{Keung:1983uu}. This generic feature of non-supersymmetric LR theories becomes somewhat subtle here. First, the heavy RH neutrino $N_1$ is tied to the extra neutralino $\widetilde Z'$ as given Eq.~(\ref{mass}), where $M_{Z'}$ is now given by Eq.~(7).
Therefore, if the Dirac mass dominates, the RH neutrino $N_1$ is not necessarily lighter than the $W_R$ gauge boson as assumed in the conventional case, and the dilepton signal would be somewhat suppressed. Second, even if $N_1$ is made lighter by taking a fairly large $m_{1/2}> M_{Z'}$, the mixing between $N_1$ and $\widetilde Z'$, equal to $M_{Z'}/m_{1/2}$ cannot be too small, for the sake of naturalness.  
In this case, one generally expects $N_1$ to decay through a slepton which could in principle be on shell (Fig.~\ref{fig1}).
Possible multi-lepton signatures from such processes are discussed separately in Ref~\cite{YZ}.

\begin{figure}[h!]
\begin{center}
\includegraphics[width=6cm]{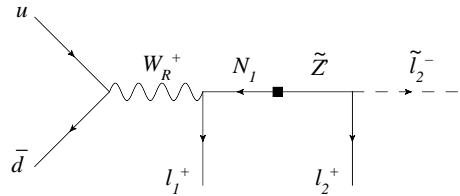}
\caption{Production and decay of the RH neutrino.}\label{fig1}
\end{center}
\end{figure}

There is a potential problem regarding the possibly low scale left-right symmetry in this model. The point is that
the scalar fields with no VEVs lead to tree-level flavor-changing processes, and must be very heavy, on the order of 10\,TeV~\cite{LRnew}.
There is a possibility of fine-tuning these effects against the usual supersymmetric ones~\cite{Zhang:2007qma}, but it goes against 
the principle of low-energy supersymmetry being behind naturalness. The trouble is that the mass of the heavy Higgs is obtained from the D-term, and thus 
roughly equal to the $W_R$ mass. The most natural outcome is then to have all the new scales raised to 10\,TeV, which would take them out of
the LHC reach. Of course, some of the soft mass terms could be made smaller as to bring some superpartner masses to the TeV region. 
In any case, barring cancellation, it is unlikely to have $W_R$ accessible at the LHC in this model. This would not change
any of our conclusions regarding the neutrino masses and our point of two light neutrinos would remain equally valid, even for very heavy $W_R$.

\subsection{Comparison with other LR models}

Before concluding this section, we briefly comment on the supersymmetric LR model with the additional Higgs fields.
It obviously makes no sense to introduce more doublets on top of the sleptons, which favors the model with triplet Higgs.
In this case, the RH scale is not automatically connected with the scale of supersymmetry breaking. 
One can use higher dimensional operators to break $SU(2)_R$ at some large scale (the AMS model)~\cite{Aulakh:1997fq}. 
The interesting feature of this theory is the necessarily exact R-parity and the stability of the LSP~\cite{Aulakh:1999cd}. 
Furthermore, one expects doubly charged super-multiplets to remain light enough to be accessible at the 
LHC~\cite{Aulakh:1997fq, Aulakh:1998nn, Chacko:1997cm, Frank:2000dw}. 

On the other hand, if the higher dimensional operators were absent or very small in this minimal model with Higgs triplets, one would be forced to break R-parity (the KM model)~\cite{Kuchimanchi:1995vk, Ji:2008cq}. Otherwise, one ends up breaking electromagnetic charge invariance. Just as in the minimal model studied here, the RH sneutrino must also develop a non-vanishing VEV, in addition to the triplet Higgs VEV.
The difference between the FS and KM models lies in the RH neutrino spectrum.
In the KM model (as in the AMS one), all the RH neutrinos are made heavy by the triplet Higgs VEVs, so it might be less appealing in view of the recent BBN and neutrino oscillation progress. As a payback, the KM model possesses richer collider phenomenology.
In the KM model, there is another contribution to the multi-lepton signal through $N_{2,3}$ and the $e^c-\widetilde W_R^+$ mixing~\cite{YZ}. Therefore, when $N_1$ is made heavier than $W_R$ (if $m_{1/2} \lesssim m_{Z'}$), such signal will be turned off in the FS model while it could still happen in the KM model. 

\section{Summary 
}

$B-L$ symmetry plays a special role in the SM: it is the only anomaly free global symmetry. 
This strongly suggests its gauging, which requires the existence of RH neutrinos, one per each family.
In turn, one predicts non-vanishing neutrino masses through the seesaw mechanism.
In the MSSM, $B-L$ symmetry can play another important role in setting the stage for the R-parity.
Recently, important progress has been made in the construction of the supersymmetric $B-L$ model, by getting rid 
of any additional Higgs beyond the MSSM ones. 
As a consequence, the $B-L$ scale is tied to the supersymmetry breaking scale, and if one believes that low energy supersymmetry is
accessible at the LHC, so is the $B-L$ gauge boson.

In this letter, we find another important characteristic feature of this theory: besides the usual actives neutrinos, one predicts
automatically two light sterile neutrinos with masses in the sub-eV region. This is the central result of our paper. It is equally valid in the left-right extension of the theory, where $B-L$ gauge symmetry follows automatically.
We further discussed some new possible phenomenological features of the LR theory relevant for LHC.

It was recently argued these new light states are favored by the BBN.  
The crucial point is that the $B-L$ gauge interactions assure them being abundant at the time of nucleosynthesis.
They also may be playing an important role in neutrino oscillations, as long as there is an oasis in the desert above
the $B-L$ scale, around $10^5\,$GeV or so.

\acknowledgements

We would like to thank Borut Bajc, Paolo Creminelli, Alejandra Melfo, Rabi Mohapatra, Miha Nemev\v{s}ek and Alexei Smirnov for useful discussions.
D.K.G. thanks ICTP High Energy Group for the hospitality where this work was
started. D.K.G. also
acknowledges partial support from the Department of Science and Technology,
India under the grant SR/S2/HEP-12/2006.
The work of G.S. and Y.Z. is partially supported by the EU FP6 Marie Curie Research and Training Network ``UniverseNetÓ (MRTN-CT-2006-035863).

\medskip
\noindent{\em Note added} \ \ After this paper was submitted to the arXiv, we learned from Rabi Mohapatra that the idea of using RH sneutrino VEV in order
to break the $SU(2)_R$ symmetry is originally due to~\cite{Hayashi:1984rd}, and in the case of $B-L$ it was used in~\cite{Mohapatra:1986aw}. Furthermore, it was noticed in~\cite{Mohapatra:1986aw} that only one RH neutrino gets a large Majorana mass which is the starting point of the analysis presented here.
This point is carefully  discussed in~\cite{Barger:2010iv}, which appeared on the same day as our paper.  We wish to acknowledge too that the
possibility of neutrinos having both Majorana and Dirac mass terms on competing level was discussed recently in the context of the 
supersymmetric $B-L$ theory in \cite{Allahverdi:2010us}. However, the neutrino spectrum and phenomenology is different from the one in this work.

\end{document}